\documentclass[11pt,a4paper]{article}

\usepackage{amsmath,amssymb}
\usepackage{graphicx}
\usepackage{booktabs}
\usepackage{multirow}
\usepackage[margin=2.5cm]{geometry}
\usepackage{xcolor}
\usepackage{listings}
\usepackage{caption}
\usepackage[colorlinks=true,linkcolor=blue,citecolor=blue,urlcolor=blue]{hyperref}

\title{HELIOS: An LLM-Driven Autonomous Indirect Trajectory Optimization Agent%
  \thanks{This work is supported by the National Natural Science Foundation of China (NSFC) under Grant No.~12572407.}}
\author{Huang An-yi\\
  \textit{State Key Laboratory of Astronautic Dynamics}}
\date{\today}

\begin{document}

\maketitle

\begin{abstract}
Low-thrust trajectory optimization is a core technology in deep-space mission design. Indirect methods based on Pontryagin's Minimum Principle (PMP) offer rigorous optimality guarantees, yet their practical application has long been hindered by three bottlenecks: (1) transversality conditions must be derived case by case for each constraint type (rendezvous, flyby, stay, gravity assist, inequality), a tedious and error-prone process; (2) different dynamics models (chemical propulsion, solar sail, $J_2$ perturbation) exhibit substantially different Hamiltonian structures, forcing repeated code rewrites; and (3) the numerical solution of shooting equations is highly sensitive to initial guesses. This paper presents HELIOS (Heuristic Engine for Low-thrust Interplanetary Optimization System), a trajectory optimization agent built around a large language model (LLM). Given a physical problem described in natural language (constraints, dynamics, optimization objective), the system autonomously performs PMP symbolic derivation, SymPy verification, C++ shooting-code generation, and numerical solution, without any human intervention.

The key innovations are: (1) a \textbf{constraint-adaptive general derivation framework} that unifies arbitrary constraints into the form $\psi(\mathbf{x}, \mathbf{p}) = 0$, automatically identifies free parameters embedded in constraints (e.g., the turning angle $\delta$ and azimuth angle $\phi$ of a gravity assist), and generates the corresponding stationarity conditions, breaking the traditional limitation of deriving transversality conditions separately for each constraint type; (2) \textbf{dynamics-adaptive four-module code generation}, in which a DYNAMICS/SWITCH/RESIDUAL/SOLVE separation architecture supports complete code generation for non-standard dynamics models (solar sail, $J_2$ perturbation, etc.) without modifying the underlying template; and (3) a \textbf{general derivation rule set (Rules G1--G4)} distilled from extensive practice, covering critical error-prone points such as $\lambda_0$ handling, control-direction verification, geometric-derivative completeness, and state-dimension consistency.

Experiments on 11 progressive test scenarios show that HELIOS correctly derives and solves problems ranging from a simple two-point rendezvous (8 variables) to multi-leg stay transfers (48 variables), gravity-assist trajectories (17 variables), and solar-sail minimum-time transfers (8 variables). The best compilation success rate reaches 100\% (11/11), and multiple scenarios converge to optimal solutions consistent with reference solutions. A multi-model comparison (8 LLM backends of different scales, total scores spanning 250--905) verifies the model-agnostic nature of the system architecture and reveals a positive correlation between model scale and derivation capability.

\end{abstract}

\section{Introduction and Related Work}

\subsection{Motivation}

Low-thrust trajectory optimization is one of the core technologies of deep-space mission design. Compared with impulsive chemical propulsion, continuous low-thrust propulsion such as electric propulsion can substantially reduce propellant consumption, at the cost of a considerably more complex trajectory optimization problem. Indirect methods, grounded in Pontryagin's Minimum Principle (PMP)~\cite{lawden1963,pontryagin1962,bryson1975}, reformulate the optimal control problem as a two-point boundary value problem (TPBVP) that is solved by shooting methods. Compared with direct methods~\cite{betts1998}, indirect methods guarantee the optimality of the solution, require no discretization, and yield a transparent solution structure (the number of bang-bang switches can be read off directly). Their practical application, however, faces the following challenges:

\textbf{Challenge 1: derivation burden caused by constraint diversity.} Different missions involve different constraint types---terminal rendezvous (position and velocity fixed), interior-point flyby (position only), stay arcs (coast arcs), gravity assists (GA), and inequality path constraints (e.g., an upper bound on the terminal velocity increment or a lower bound on the periapsis radius). The transversality conditions of each constraint type take a different form and must be derived one by one. A gravity-assist constraint, for example, involves costate jump conditions at the flyby epoch together with a Hamiltonian continuity condition, and the derivation requires partial derivatives of the constraint functions with respect to both the state variables and free parameters (turning angle, periapsis radius), making it highly error-prone.

\textbf{Challenge 2: code rewriting caused by dynamics diversity.} Different spacecraft correspond to different dynamics models---chemical/electric propulsion (with a mass state and bang-bang control), solar sail (no mass state, continuous cone-angle control), $J_2$ perturbation (non-spherical gravity field), and so on. The Hamiltonian, costate equations, and optimal control law differ substantially across models, and the numerical code must be written and validated separately for each of them.

\textbf{Challenge 3: numerical convergence of the shooting equations.} The shooting system arising from an indirect formulation is highly nonlinear, and the probability that a random initial guess falls inside the basin of convergence decays exponentially with the number of variables. Once the variable count exceeds 20, simple random-restart strategies often fail.

In recent years, large language models (LLMs) have demonstrated strong capabilities in symbolic reasoning and code generation. This paper presents HELIOS (Heuristic Engine for Low-thrust Interplanetary Optimization System), a trajectory optimization agent with an LLM at its core. The central design principle is: \emph{the problem description states only physical facts and contains no PMP derivation results; starting from the primitive form of the minimum principle, the LLM autonomously completes the entire derivation and code generation.} This design endows the system with general adaptability across constraint types, dynamics models, and optimization objectives.

\subsection{Related Work}

\textbf{Low-thrust indirect methods.} Indirect low-thrust trajectory optimization has been extensively studied. The Sims--Flanagan method~\cite{sims1999} handles the discontinuity of bang-bang control through shooting formulations with switching functions. Russell~\cite{russell2007} applied primer vector theory to global low-thrust trade studies. Homotopy-based smoothing strategies ($\varepsilon$-homotopy)~\cite{bertrand2002,haberkorn2004} significantly improve convergence by continuously deforming an easily solved smooth problem into the original bang-bang problem; practical homotopic techniques for low-thrust problems are summarized by Jiang et al.~\cite{jiang2012}. These methods form the numerical foundation of the HELIOS solving framework. On the global-optimization side, shape-based approximations~\cite{petropoulos2004,taheri2012}, solar-electric-propulsion mission design~\cite{casalino2011}, complex interplanetary trajectory design in the GTOC context~\cite{izzo2016}, and inflationary differential evolution~\cite{vasile2011} provide complementary global search capabilities. Solar-sail dynamics and mission applications are treated comprehensively by McInnes~\cite{mcinnes1999}.

\textbf{LLMs for scientific computing.} Large language models have shown remarkable few-shot learning~\cite{brown2020} and chain-of-thought reasoning~\cite{wei2022} abilities, and models trained on code~\cite{chen2021} can synthesize nontrivial programs. Frontier general-purpose models~\cite{achiam2023,liu2024} further extend these capabilities to multi-step symbolic reasoning. In aerospace dynamics, prior work has explored LLM-assisted orbit design, mission planning, and fault diagnosis. To the best of our knowledge, however, using an LLM for the end-to-end generation of PMP symbolic derivations and shooting code has not been reported.

\textbf{Prompt engineering.} The derivation capability of HELIOS originates largely from a set of general rules embedded in the prompt. These rules are not problem-specific ``cheat'' hints but general knowledge distilled from PMP theory, consistent in spirit with chain-of-thought prompting~\cite{wei2022}: they decompose a complex symbolic derivation into verifiable intermediate steps.

\subsection{Contributions}

\begin{enumerate}
  \item \textbf{Constraint-adaptive derivation framework.} Arbitrary constraints are unified into the form $\psi(\mathbf{x}, \mathbf{p}) = 0$; free parameters embedded in constraints (Category~C) are automatically identified, and the corresponding stationarity conditions $\partial(\sum_k \nu_k \psi_k)/\partial p = 0$ are generated. This removes the traditional limitation of deriving transversality conditions separately for each constraint type.
  \item \textbf{Dynamics-adaptive code generation.} A four-module DYNAMICS/SWITCH/RESIDUAL/SOLVE separation architecture supports complete code generation for non-standard dynamics models. For standard low-thrust problems only two modules (RESIDUAL + SOLVE) need to be generated; for solar-sail-type models all four modules are generated.
  \item \textbf{General derivation rule set.} Rules G1--G4 cover $\lambda_0$ handling, control-direction verification, geometric-derivative completeness, and state-dimension consistency; Rule~B2 handles free parameters inside constraints. These rules are general knowledge distilled from extensive practice and are not tied to any specific problem.
  \item \textbf{Multi-model backend validation.} The model-agnostic nature of the system architecture is verified on 8 LLMs of different scales, with total scores spanning 250--905, providing experimental evidence for backend selection. Token consumption of each model (465k--836k) is also reported as a reference for deployment-cost estimation.
\end{enumerate}

\section{System Architecture}

\subsection{Overall Design}

\begin{figure}[htbp]
  \centering
  \includegraphics[width=0.60\textwidth]{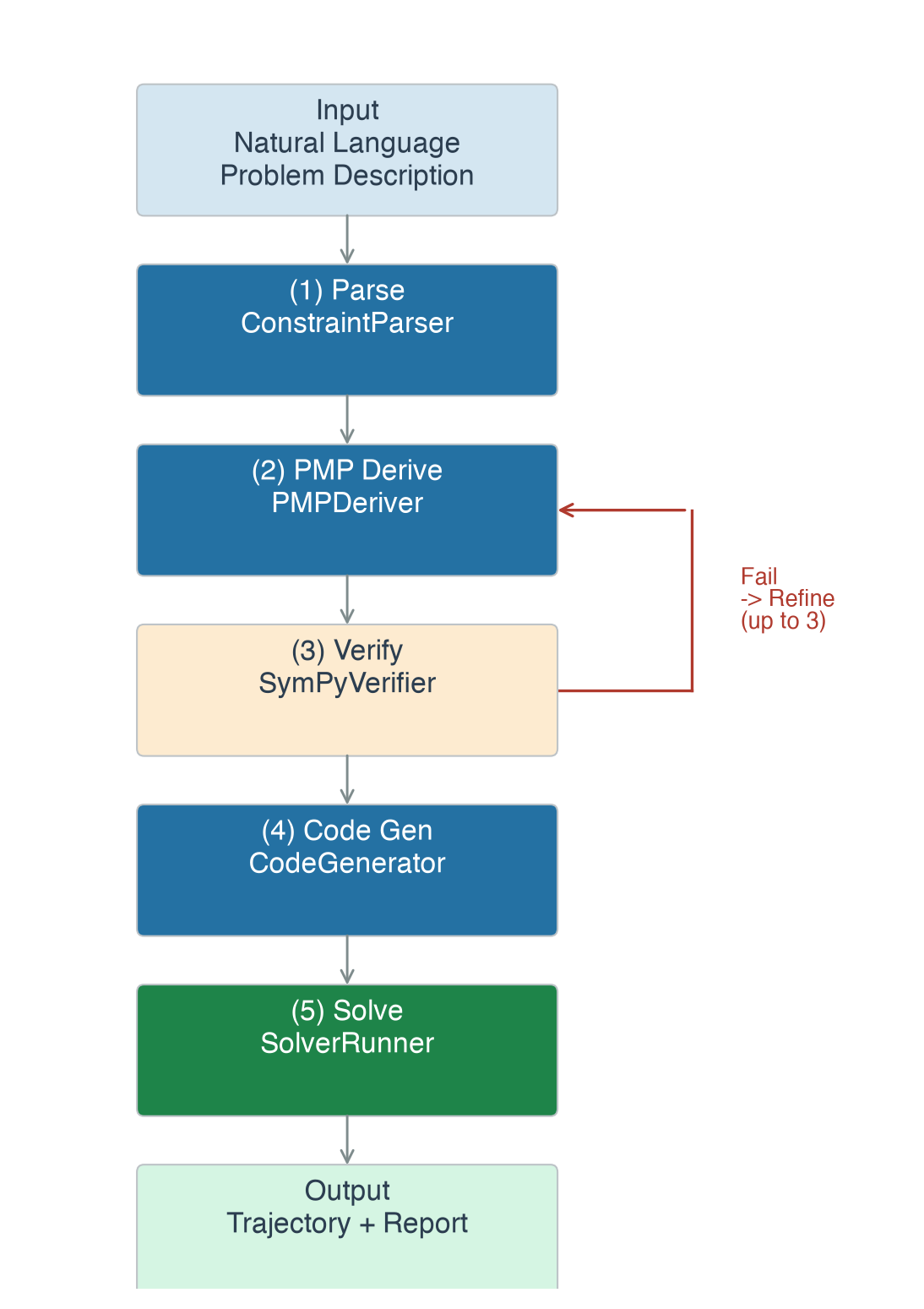}
  \caption{HELIOS five-stage pipeline architecture. The natural-language problem description is parsed into a structured specification, from which the PMP derivation, symbolic verification, code generation, and numerical solution are produced autonomously. Failed verification or compilation triggers bounded refinement loops.}
  \label{fig:arch}
\end{figure}

HELIOS adopts a five-stage pipeline architecture (Figure~\ref{fig:arch}). The user describes the physical problem in natural language (target body, constraint types, dynamics model, spacecraft parameters, optimization objective), and the system sequentially performs constraint parsing, PMP derivation, symbolic verification, code generation, and numerical solving.

\begin{lstlisting}[caption={Five-stage pipeline of HELIOS.},label={lst:pipeline}]
User input (natural-language problem description)
  |
  v
Stage 1: Constraint parsing (ConstraintParser)
  |  natural language -> structured JSON (problem_spec)
  v
Stage 2: PMP derivation (PMPDeriver)
  |  problem_spec -> costate equations + transversality conditions
  |                   + shooting variables/equations
  |  (refinement loop on verification failure, up to 3 rounds)
  v
Stage 3: SymPy symbolic verification (SymPyVerifier)
  |  correctness of costate ODEs -> executed as a Python symbolic script
  v
Stage 4: Code generation (CodeGenerator)
  |  PMP results + problem_spec -> C++ shooting code (2 or 4 modules)
  v
Stage 5: Compilation and solving (SolverRunner)
  |  g++ compilation + cminpack hybrd1_ solving + result parsing
  v
Output: trajectory, dv, mass history, convergence report
\end{lstlisting}

\subsection{Constraint Parser (Stage 1)}

The ConstraintParser converts the natural-language problem into a structured JSON object (\texttt{problem\_spec}) containing: the central body and its gravitational constant; departure/arrival states (Cartesian coordinates or orbital elements); spacecraft parameters (thrust, specific impulse, mass, or solar-sail characteristic acceleration); a description of the dynamics model; the constraint list (terminal, interior-point, path); and the optimization objective (minimum fuel / minimum time).

A key design choice is that the parser does not distinguish among constraint types---all constraints are uniformly represented as \texttt{constraint} objects carrying position, time, type, and target information. The LLM decides how each constraint is treated during the PMP derivation stage.

\subsection{PMP Derivation (Stage 2)}

The PMPDeriver is the core intelligent module of the system. Given the \texttt{problem\_spec} and the PMP derivation prompt (\texttt{pmp\_derive.txt}), the LLM outputs the complete indirect shooting formulation:

\begin{itemize}
  \item the Hamiltonian $H$ (with an explicit sign convention: min/max principle);
  \item the costate differential equations
  \begin{equation}
    \dot{\boldsymbol{\lambda}} = -\frac{\partial H}{\partial \mathbf{x}};
    \label{eq:costate}
  \end{equation}
  \item the optimal control law (a bang-bang switching function or a continuous control formula);
  \item all transversality conditions (terminal + interior-point + free-time);
  \item the list of shooting variables (with the physical meaning and initialization strategy of each);
  \item the list of shooting equations (in one-to-one correspondence with the variables).
\end{itemize}

The general Hamiltonian under the minimum-principle convention is
\begin{equation}
  H = \lambda_0 \cdot L + \boldsymbol{\lambda}^T \cdot f(\mathbf{x}, u, t),
  \label{eq:hamiltonian}
\end{equation}
where $\lambda_0$ is the abnormal multiplier, $L$ is the running cost, $\boldsymbol{\lambda}$ is the costate vector, and $f(\mathbf{x}, u, t)$ is the dynamics right-hand side. The core rules embedded in the prompt include:

\textbf{Rule 0 (constraints are equations).} Every constraint appearing in the problem statement must appear as one of the shooting equations. Terminal constraints, interior-point constraints, and boundary conditions alike must be present in the shooting system.

\textbf{Rule B (interior-point constraints).} Each interior-point constraint $\psi(\mathbf{x}(t_k)) = 0$ introduces a multiplier $\boldsymbol{\nu}$, and the costate jumps at $t_k$ according to
\begin{equation}
  \boldsymbol{\lambda}^{+}(t_k) = \boldsymbol{\lambda}^{-}(t_k) + \boldsymbol{\nu}^T \frac{\partial \psi}{\partial \mathbf{x}}.
  \label{eq:jump}
\end{equation}
The multiplier $\boldsymbol{\nu}$ is added as a shooting variable, and the constraint $\psi = 0$ is added as a shooting equation.

\textbf{Rule B2 (constraint parameters).} Scan the expression of every constraint $\psi$ and identify free parameters $p$ that are neither state variables nor multipliers (e.g., the GA turning angle $\delta$, azimuth angle $\phi$, and periapsis radius $r_p$). Each such parameter adds one shooting variable and one stationarity equation
\begin{equation}
  \sum_k \nu_k \, \frac{\partial \psi_k}{\partial p} = 0.
  \label{eq:stationarity}
\end{equation}
This rule enables the system to automatically handle problems with parameterized constraints such as gravity assists.

\textbf{Rule G1 ($\lambda_0$ handling).} For problems with a cost functional, $\lambda_0$ must be a shooting variable and must be included in the spherical normalization.

\textbf{Rule G2 (control-direction verification).} The optimal control direction must be verified to maximize (or minimize) $H$. The sign convention ($+s$ or $-s$) depends on whether the max or min principle is adopted.

\textbf{Rule G3 (geometric derivatives).} When the optimal control direction $\mathbf{n}$ depends on the state $\mathbf{x}$ (e.g., the sail normal depends on $\hat{\mathbf{r}}$), the costate equations must include the geometric-derivative terms $\partial \mathbf{n}/\partial \mathbf{x}$. The envelope theorem eliminates only $\partial H/\partial \alpha$ (the control variable itself), not $\partial \mathbf{n}/\partial \mathbf{x}$.

\textbf{Rule G4 (state dimension).} Variables that do not appear in the dynamics equations (e.g., the mass of a solar sail) are not state variables and must be excluded.

\subsection{Symbolic Verification (Stage 3)}

\begin{figure}[htbp]
  \centering
  \includegraphics[width=0.75\textwidth]{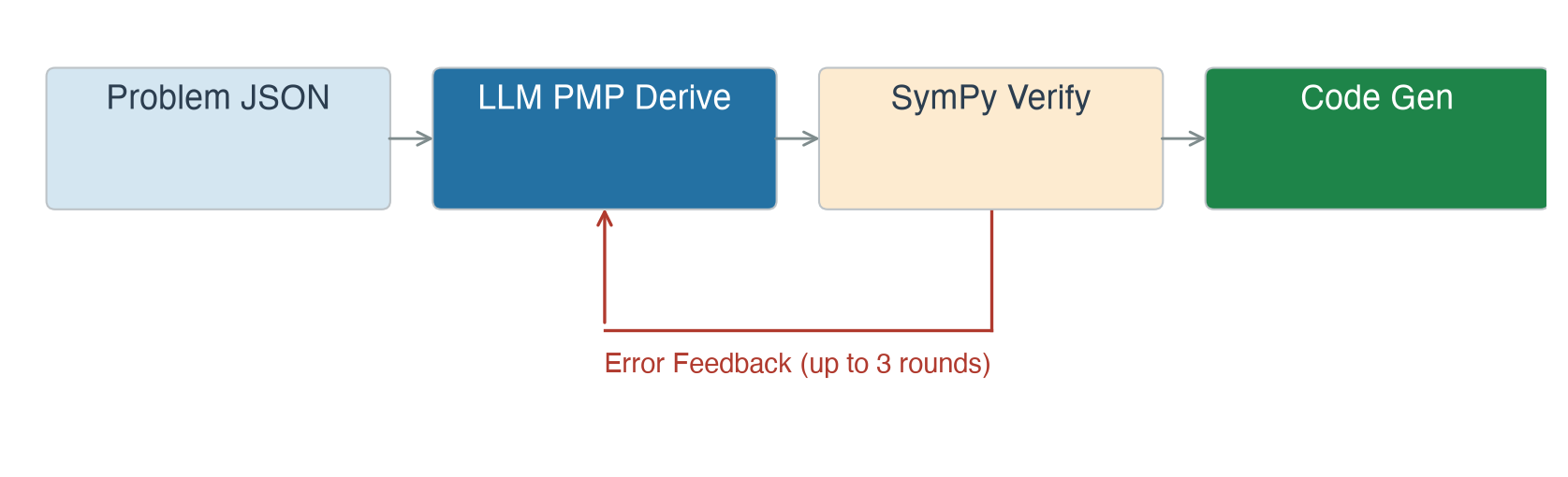}
  \caption{PMP derivation and SymPy verification loop. The derived costate equations are checked by symbolic differentiation; failures are fed back to the deriver for up to three refinement rounds.}
  \label{fig:verify}
\end{figure}

The SymPyVerifier executes the derived costate ODEs as a Python symbolic script, computes $-\partial H/\partial \mathbf{x}$ symbolically, and compares it against the LLM's derivation (Figure~\ref{fig:verify}). On verification failure, the error message is fed back to the PMPDeriver for refinement (up to 3 rounds).

For complex dynamics models (e.g., solar sail), the symbolic derivation involves deeply nested chain-rule expansions, and SymPy's \texttt{simplify} may time out. In that case the system skips verification and proceeds directly to code generation, relying on the correctness of the LLM's derivation.

\subsection{Code Generation (Stage 4)}

The CodeGenerator translates the PMP derivation results into C++ shooting code based on a fixed framework template. The template is divided into a fixed part and an LLM-generated part:

\textbf{Fixed part:} normalization-constant initialization, orbital mechanics (Kepler propagation / fixed Cartesian states), an RK4 integrator, an adaptive $\varepsilon$-homotopy solver, and the MINPACK \texttt{hybrd1\_} interface.

\textbf{LLM-generated part} (2 or 4 modules, see the table below):

\begin{table}[htbp]
  \centering
  \caption*{Module generation strategy of the CodeGenerator.}
  \begin{tabular}{@{}lll@{}}
    \toprule
    Module & Standard low-thrust & Non-standard dynamics \\
    \midrule
    DYNAMICS & Template default (\texttt{fmax}/\texttt{fmid}/\texttt{fnull}) & LLM-generated custom RHS \\
    SWITCH   & Template default (bang-bang switching) & LLM-generated control law \\
    RESIDUAL & LLM-generated (\texttt{objective\_RK}) & LLM-generated \\
    SOLVE    & LLM-generated (\texttt{main} + initial guess) & LLM-generated \\
    \bottomrule
  \end{tabular}
\end{table}

For standard low-thrust problems (chemical/electric propulsion, bang-bang control), only the RESIDUAL and SOLVE modules are generated, while DYNAMICS and SWITCH use the template defaults. For non-standard dynamics such as solar sails, all four modules must be generated---the LLM writes the right-hand-side function itself (e.g., the solar-sail acceleration), the control law (the optimal cone-angle formula), and the propagation logic.

\textbf{Ephemeris support.} The \texttt{compute\_target\_rv} function supports two modes---Keplerian propagation from orbital elements (\texttt{eph[0] >= 0}) and direct assignment of fixed Cartesian states (\texttt{eph[0] == -1}). The latter is used in scenarios where the problem provides exact positions and velocities that require no propagation.

\subsection{Numerical Solving (Stage 5)}

The generated C++ code is compiled with g++ and linked against the cminpack library. The solution procedure adopts an adaptive $\varepsilon$-homotopy strategy:

\begin{enumerate}
  \item \textbf{Hot start ($\varepsilon = 1.0$).} Under the fully smoothed continuous-thrust problem, multiple random restarts are performed (each invoking \texttt{hybrd1\_}), and the initial guess with the smallest residual is retained.
  \item \textbf{$\varepsilon$ descent.} If the current $\varepsilon$ converges, the step size \texttt{dete} is decreased geometrically ($0.05 \to 0.01 \to 0.001$); if it fails to converge, a half-step retreat is attempted.
  \item \textbf{Termination.} $\varepsilon \le \texttt{minepsl}$ (typically 0.0).
\end{enumerate}

\section{Key Methods}

\subsection{Unified Constraint Framework}

The core innovation of HELIOS is the unification of all constraints into the form
\begin{equation}
  \psi(\mathbf{x}, p_1, p_2, \ldots) = 0,
  \label{eq:constraint}
\end{equation}
where $\mathbf{x}$ denotes the state variables and $p$ the free parameters embedded in the constraint. The processing pipeline is:

\begin{enumerate}
  \item \textbf{Scan.} List all constraints of the problem (terminal, interior-point, path, boundary).
  \item \textbf{Classify.} Classify the variables in each constraint into state variables (Category~A), multipliers (Category~B), and free parameters (Category~C).
  \item \textbf{Shooting variables.} State variables require no additional variables; the multiplier $\boldsymbol{\nu}$ of each interior-point constraint becomes a shooting variable (Rule~B); each free parameter $p$ becomes a shooting variable (Rule~B2).
  \item \textbf{Shooting equations.} Each constraint $\psi = 0$ is itself a shooting equation; each free parameter $p$ generates the stationarity equation \eqref{eq:stationarity}; each free time $t_k$ generates the transversality condition
  \begin{equation}
    H(t_k^{-}) - H(t_k^{+}) - \sum \nu \, \frac{\partial \psi}{\partial t_k} = 0.
    \label{eq:transversality}
  \end{equation}
\end{enumerate}

This unified treatment allows the system to adapt automatically to arbitrary constraint combinations. Taking the gravity assist as an example:

\begin{lstlisting}[caption={Gravity-assist constraint structure handled automatically by the unified framework.},label={lst:ga}]
Constraint 1: r(tm-) - r_mars(tm) = 0               (3 eqs, multipliers nu_r(3))
Constraint 2: v(tm+) - v_mars(tm) - R(delta,phi)*v_inf- = 0
                                                   (3 eqs, multipliers nu_v(3))
Free parameters: delta, phi (Category C)            (2 vars, 2 stationarity eqs)
Free time:       tm                                 (1 var, 1 transversality eq)
Inequality:      rp - rmin >= 0                     (two-pass method)
\end{lstlisting}

The system automatically identifies $\delta$ and $\phi$ as Category~C parameters and generates the shooting variables together with the stationarity equations $\boldsymbol{\nu}_v \cdot \partial R/\partial \delta \cdot \mathbf{v}_{\infty}^{-} = 0$ and $\boldsymbol{\nu}_v \cdot \partial R/\partial \phi \cdot \mathbf{v}_{\infty}^{-} = 0$. No manual derivation of GA-specific transversality conditions is required.

\subsection{Dynamics Model Adaptation}

The essential differences among dynamics models lie in the state dimension, the control parameterization, and the Hamiltonian structure. HELIOS handles them through the following mechanisms:

\textbf{Standard low-thrust} (chemical/electric propulsion): a 14-dimensional state (7 states + 7 costates), bang-bang control, using the template defaults \texttt{fmax}/\texttt{fmid}/\texttt{fnull} and the \texttt{md14} switch-detection integrator.

\textbf{Solar sail:} a 12-dimensional state (6 states + 6 costates, no mass), continuous cone-angle control, with an LLM-generated custom RHS and optimal cone-angle formula. The control law is
\begin{equation}
  \mathbf{n} = \cos\alpha \, \hat{\mathbf{r}} + \sin\alpha \, \hat{\mathbf{s}},
  \label{eq:sailnormal}
\end{equation}
with the optimal cone angle
\begin{equation}
  \alpha^{*} = \arctan\!\left( \frac{-3a + \sqrt{9a^{2} + 8b^{2}}}{4b} \right),
  \qquad a = \boldsymbol{\lambda}_v \cdot \hat{\mathbf{r}},\quad b = \lVert \boldsymbol{\lambda}_{v\perp} \rVert,
  \label{eq:coneangle}
\end{equation}
where $\hat{\mathbf{s}} = \boldsymbol{\lambda}_{v\perp} / \lVert \boldsymbol{\lambda}_{v\perp} \rVert$ and $\boldsymbol{\lambda}_{v\perp}$ is the component of the velocity costate perpendicular to $\hat{\mathbf{r}}$.

\textbf{$J_2$ perturbation:} a $J_2$ acceleration term is added on top of the standard low-thrust model, and the costate equations include the partial derivatives of the $J_2$ acceleration with respect to position.

The four-module code-generation architecture allows the LLM to produce complete custom code for each dynamics model while retaining the template's numerical infrastructure (normalization, integrator, solver).

\subsection{Numerical Stability Strategies}

\textbf{Normalization.} All physical quantities are transformed into normalized units (AU, year, AU/year). For heliocentric problems, $\mu \approx 39.48$, positions are ${\sim}2$~AU, velocities ${\sim}5$~AU/yr, and all variables lie in the $O(1)$--$O(10)$ range.

\textbf{Time-fraction parameterization.} Free time variables are stored as fractions $\tau \in [0, 1]$, avoiding the magnitude mismatch between absolute MJD values (${\sim}60000$) and costate variables (${\sim}O(1)$).

\textbf{Target-epoch evaluation.} The target body's state is evaluated at the actual arrival epoch \texttt{T\_initial + tau*tf\_day}, rather than at a fixed epoch.

\textbf{Dimensionless residuals.} \texttt{fvec} uses only raw normalized differences without any secondary scaling (avoiding false convergence caused by, e.g., dividing by \texttt{R\_unit}).

\textbf{Sign convention of $H$.} The min principle ($H = \lambda_0 L + \boldsymbol{\lambda}^T f$) and the max principle ($H = -\lambda_0 L + \boldsymbol{\lambda}^T f$) are declared explicitly in the derivation to avoid sign confusion.

\section{Experimental Design}

\subsection{Test Scenarios}

Eleven progressive test scenarios were designed, covering different constraint types, dynamics models, and optimization objectives (see the table below).

\begin{table}[htbp]
  \centering
  \caption*{Eleven progressive test scenarios. The Source column lists the reference from which each scenario is drawn: S01--S04 and S07 from~\cite{huang2015}; S05 and S06 from~\cite{huang2018}; S08 and S09 from~\cite{huang2016}; S10 from~\cite{jiang2012}; S11 from~\cite{sun2019}.}
  \small
  \resizebox{\textwidth}{!}{%
  \begin{tabular}{@{}cllllclc@{}}
    \toprule
    \# & Scenario & Constraint type & Dynamics & Objective & Vars & Difficulty & Source \\
    \midrule
    01 & Fixed-time rendezvous       & Terminal rendezvous              & Two-body  & Min fuel          & 8  & L1 & \cite{huang2015} \\
    02 & Min-time rendezvous        & Free final time                  & Two-body  & Min time          & 9  & L2 & \cite{huang2015} \\
    03 & Multi interior rendezvous  & 3 interior rendezvous            & Two-body  & Min fuel          & 29 & L3 & \cite{huang2015} \\
    04 & Interior flyby             & Interior flyby                   & Two-body  & Min fuel          & 12 & L2 & \cite{huang2015} \\
    05 & $J_2$-perturbed transfer   & Terminal rendezvous              & Two-body + $J_2$ & Min fuel     & 8  & L2 & \cite{huang2018} \\
    06 & $J_2$ LEO rendezvous       & Terminal rendezvous              & Two-body + $J_2$ & Min fuel     & 8  & L2 & \cite{huang2018} \\
    07 & Stay + $t_0$ sliding       & $3\times$STAY + free $t_0$       & Two-body  & Min fuel          & 48 & L4 & \cite{huang2015} \\
    08 & Stay + mass drop + flyby  & $2\times$STAY + mass drop + flyby & Two-body & Min initial mass & 22 & L3 & \cite{huang2016} \\
    09 & Stay + velocity inequality & $2\times$STAY + inequality      & Two-body  & Min initial mass  & 23 & L4 & \cite{huang2016} \\
    10 & Mars gravity assist        & Gravity assist                   & Two-body  & Min fuel          & 17 & L3 & \cite{jiang2012} \\
    11 & Solar-sail min-time        & Terminal rendezvous              & Two-body (sail) & Min time   & 8  & L2 & \cite{sun2019} \\
    \bottomrule
  \end{tabular}}%
\end{table}

The scenarios span a complete difficulty gradient from a simple 8-variable problem to a complex 48-variable one. Constraint types include terminal rendezvous, interior flyby, interior rendezvous, STAY (stay arcs), mass drop, gravity assist, and inequality constraints. Dynamics models include two-body point-mass gravity (with bang-bang thrust), two-body with $J_2$ perturbation, and two-body solar sail (continuous control, no mass state).

\subsection{Evaluation Metrics}

A five-dimension scoring system is adopted (100 points per scenario, 1100 points in total), as shown in the table below.

\begin{table}[htbp]
  \centering
  \caption*{Five-dimension scoring system.}
  \begin{tabular}{@{}llp{8.5cm}@{}}
    \toprule
    Dimension & Full score & Evaluation content \\
    \midrule
    d1 Parsing      & 30 & Whether constraints, dynamics, and parameters are correctly extracted \\
    d2 Derivation   & 20 & Hamiltonian, costate ODEs, optimal control law, variable/equation counts \\
    d3 Compilation  & 20 & Complete C++ structure, successful compilation \\
    d4 Convergence  & 20 & Shooting residual $\max|\texttt{fvec}| < 10^{-5}$ \\
    d5 Accuracy     & 10 & Deviation of $m_f$/$\Delta v$ from the reference solution \\
    \bottomrule
  \end{tabular}
\end{table}

\subsection{Multi-Model Baselines}

The same test scenarios were run on 8 LLMs of different scales (see the table below) to evaluate the model-agnostic nature of the system architecture.

\begin{table}[htbp]
  \centering
  \caption*{LLM backends used in the evaluation.}
  \begin{tabular}{@{}ll@{}}
    \toprule
    Scale & Models \\
    \midrule
    Large  & DeepSeek-V4-Pro, Qwen3.7-Max, GLM-5.2 \\
    Medium & Qwen3.6-27B, DeepSeek-V3.2-Instruct, MiniMax-M3, DeepSeek-V4-Flash \\
    Small  & GLM-4-9B \\
    \bottomrule
  \end{tabular}
\end{table}

All models use identical prompts, temperature (0.1), and pipeline configuration, with thinking mode disabled. The per-scenario timeout is 1200 seconds, and the compilation-repair loop runs at most 3 rounds.

\textbf{Model selection rationale.} The 8 LLM backends are selected from open-source (open-weight) model families---DeepSeek (DeepSeek-AI), Qwen (Alibaba), GLM (Zhipu AI), and MiniMax---covering large, medium, and small parameter scales. Open-source models are chosen primarily for \emph{reproducibility}: their weights and version identifiers are publicly available, allowing other researchers to replicate the evaluation under identical configurations. In addition, the selected models span a wide capability range (total scores 250--905), enabling the evaluation to reveal the relationship between model scale and derivation capability---the primary interest of this study---rather than identifying the single best LLM. Extending the benchmark to proprietary models such as GPT-5 and Claude is straightforward given the model-agnostic pipeline design and is left as future work.

\section{Experimental Results}

\subsection{Multi-Model Comparison}

\begin{figure}[htbp]
  \centering
  \includegraphics[width=0.85\textwidth]{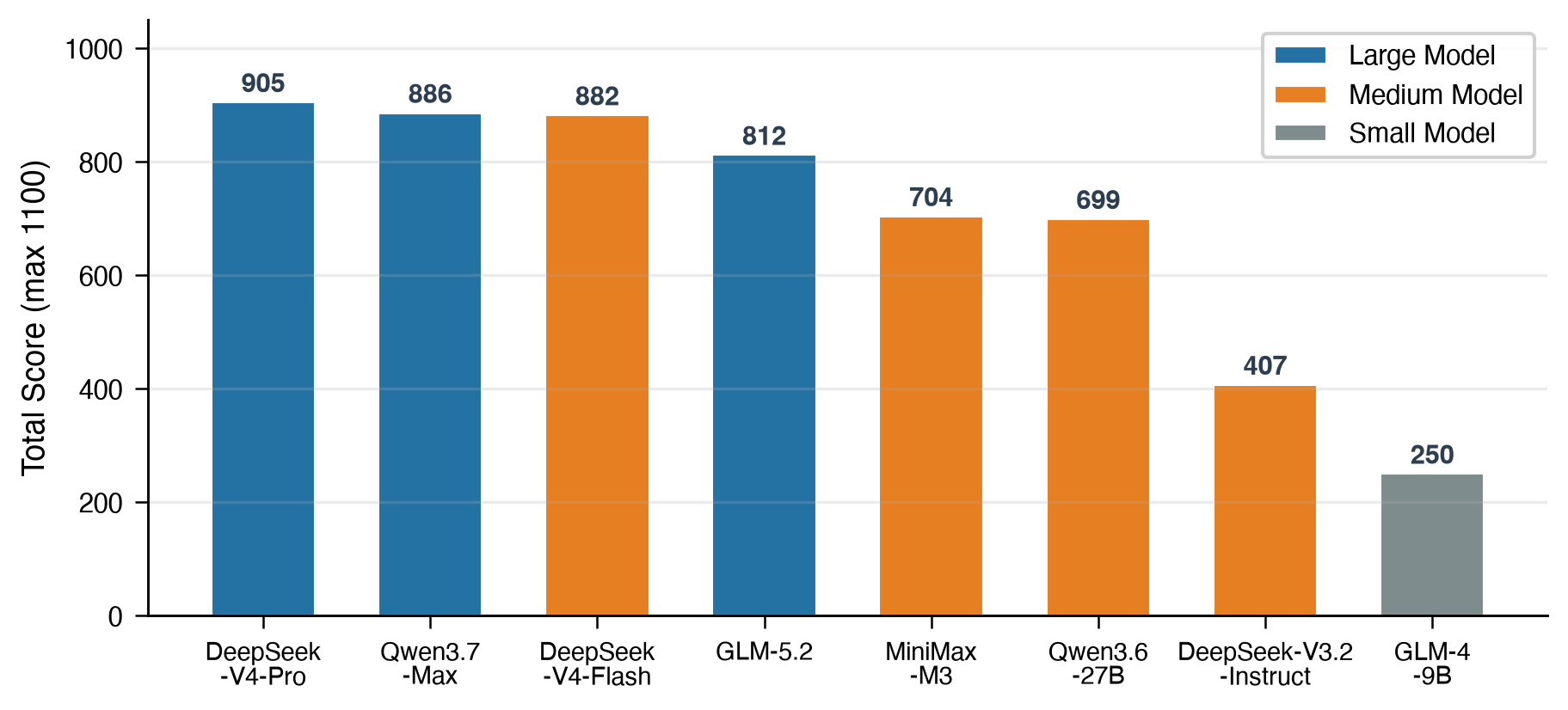}
  \caption{Total scores of the 8 LLM backends across the 11 scenarios (full score 1100).}
  \label{fig:comparison}
\end{figure}

Table~\ref{tab:comparison} presents the composite scores of the 8 LLM backends across the 11 scenarios, and Figure~\ref{fig:comparison} visualizes the total scores.

\begin{table}[htbp]
  \centering
  \caption{Multi-model evaluation results (full score 1100). ``Conv.'' and ``Comp.'' count the scenarios that converged and compiled, respectively.}
  \label{tab:comparison}
  \resizebox{\textwidth}{!}{%
  \begin{tabular}{@{}cllcccccccccccccc@{}}
    \toprule
    Rank & Model & Scale & S01 & S02 & S03 & S04 & S05 & S06 & S07 & S08 & S09 & S10 & S11 & Total & Conv. & Comp. \\
    \midrule
    1 & DeepSeek-V4-Pro       & Large  & 100 & 85 & 75 & 80 & 70 & 70 & 100 & 100 & 100 & 70 & 55 & \textbf{905} & 4 & 11 \\
    2 & Qwen3.7-Max           & Large  & 100 & 97 & 70 & 80 & 70 & 70 & 20  & 100 & 100 & 97 & 82 & \textbf{886} & 6 & 10 \\
    3 & DeepSeek-V4-Flash     & Medium & 100 & 70 & 70 & 80 & 70 & 70 & 100 & 100 & 70  & 70 & 82 & \textbf{882} & 4 & 11 \\
    4 & GLM-5.2               & Large  & 100 & 75 & 70 & 80 & 70 & 70 & 70  & 100 & 20  & 97 & 60 & \textbf{812} & 3 & 10 \\
    5 & MiniMax-M3            & Medium & 100 & 97 & 80 & 80 & 70 & 50 & 20  & 97  & 20  & 70 & 20 & \textbf{704} & 3 & 7  \\
    6 & Qwen3.6-27B           & Medium & 100 & 97 & 20 & 80 & 70 & 70 & 20  & 70  & 20  & 70 & 82 & \textbf{699} & 3 & 8  \\
    7 & DeepSeek-V3.2-Instruct & Medium & 80 & 97 & 20 & 20 & 70 & 20 & 20  & 20  & 20  & 20 & 20 & \textbf{407} & 1 & 3  \\
    8 & GLM-4-9B              & Small  & 20  & 20 & 20 & 20 & 20 & 50 & 20  & 20  & 20  & 20 & 20 & \textbf{250} & 0 & 0  \\
    \bottomrule
  \end{tabular}}
\end{table}

\textbf{Key findings:}

\begin{enumerate}
  \item \textbf{DeepSeek-V4-Pro achieves the highest total score (905/1100)}, converging on complex scenarios such as S07 (48-variable STAY) and S09 (inequality constraints), demonstrating strong derivation capability.
  \item \textbf{Qwen3.7-Max converges on the most scenarios (6/11)}, with outstanding performance on S10 (Mars gravity assist) and S11 (solar sail).
  \item \textbf{DeepSeek-V4-Flash achieves the highest compilation rate (11/11)}, generating compilable code for every scenario.
  \item \textbf{Total score correlates positively with model scale}: large models average 868 points, medium models 698, and the small model 250.
  \item \textbf{GLM-4-9B (9B) fails to complete the task}, indicating a minimum capability threshold for trajectory-optimization derivation.
\end{enumerate}

\subsection{Token Consumption}

Table~\ref{tab:token} reports the token consumption of each model (estimated from the character counts of the LLM logs).

\begin{table}[htbp]
  \centering
  \caption{Token consumption statistics (estimated values).}
  \label{tab:token}
  \begin{tabular}{@{}llcccc@{}}
    \toprule
    Model & Scale & Total input & Total output & Total & Per-scenario avg. \\
    \midrule
    Qwen3.6-27B            & Medium & 571k & 264k & 836k & 76k \\
    MiniMax-M3             & Medium & 532k & 301k & 834k & 76k \\
    DeepSeek-V4-Pro        & Large  & 526k & 268k & 794k & 72k \\
    Qwen3.7-Max            & Large  & 550k & 224k & 774k & 70k \\
    DeepSeek-V4-Flash      & Medium & 539k & 223k & 761k & 69k \\
    GLM-5.2                & Large  & 514k & 179k & 693k & 63k \\
    GLM-4-9B               & Small  & 523k & 131k & 653k & 59k \\
    DeepSeek-V3.2-Instruct & Medium & 329k & 136k & 465k & 42k \\
    \bottomrule
  \end{tabular}
\end{table}

Input-token consumption is essentially identical across models (${\sim}530$k), since the prompt templates are the same. The differences come mainly from output tokens---models that undergo more refinement rounds produce longer outputs. DeepSeek-V3.2-Instruct consumes the least overall (465k), because most of its scenarios never reach the code-generation stage.

\subsection{Derivation Quality Analysis}

The agent performs correctly on the following critical derivation points:

\begin{enumerate}
  \item \textbf{Sign of transversality conditions:} all scenarios output $H(t_k^{-}) - H(t_k^{+}) - \boldsymbol{\nu}\cdot\mathbf{v}_{\text{target}} - \boldsymbol{\nu}_v\cdot\mathbf{a}_{\text{target}} = 0$ without sign errors.
  \item \textbf{$\lambda_0$ handling:} both min-time and min-fuel scenarios correctly include $\lambda_0$ in the spherical normalization.
  \item \textbf{State dimension:} the solar-sail scenario correctly excludes the mass state (Rule~G4), using a 12-dimensional state.
  \item \textbf{Constraint-parameter identification:} the GA scenario correctly identifies $\delta$ and $\phi$ as shooting variables (Rule~B2).
  \item \textbf{Geometric derivatives:} the solar-sail scenario includes the $\sin(\alpha - \varphi)\,\hat{\mathbf{s}}$ term (Rule~G3).
  \item \textbf{STAY multiple shooting:} S07 (48 variables) correctly implements arrival/departure costate separation, and the $H$-jump contains both endpoint terms.
  \item \textbf{$J_2$ costate gradient:} S05/S06 correctly derive the full $3\times3$ Jacobian of the $J_2$ acceleration (Rule~G6).
  \item \textbf{Solar-sail control law:} S11 correctly uses the cubic equation corresponding to $\cos^2\alpha$, rather than an \texttt{atan2} form.
\end{enumerate}

\subsection{Compilation Repair Loop Effectiveness}

The compilation repair loop (at most 3 rounds) substantially improves the compilation success rate (see the table below).

\begin{table}[htbp]
  \centering
  \caption*{Effect of the compilation repair loop on compilation success (number of scenarios compiled, out of 11).}
  \begin{tabular}{@{}lccc@{}}
    \toprule
    Model & Before repair & After repair & Improvement \\
    \midrule
    DeepSeek-V4-Flash & 4 & 11 & +7 \\
    GLM-5.2           & 4 & 10 & +6 \\
    MiniMax-M3        & 1 & 7  & +6 \\
    Qwen3.6-27B       & 3 & 8  & +5 \\
    \bottomrule
  \end{tabular}
\end{table}

Common repair types include duplicated variable definitions (removing the second declaration), function-signature mismatches (wrong number of \texttt{dot3} arguments), and undeclared identifiers (adding forward declarations). The repair loop only fixes compilation errors and never alters the mathematical logic.

\subsection{Case Study: S07 Stay Constraints + $t_0$ Sliding (48 Variables)}

To demonstrate the end-to-end autonomous derivation capability of HELIOS, this section walks through the inputs and outputs of the five-stage pipeline on scenario S07. S07 has the largest variable count (48) and the most complex constraint structure among the 11 scenarios, comprising 3 interior rendezvous constraints, 3 stay-arc (coast) constraints, and 1 free departure-epoch constraint.

\textbf{Problem statement.} The spacecraft departs from asteroid Azusa and rendezvous sequentially with the four asteroids Nikko, Magri, Tengstrom, and SIMBAD. After each interior rendezvous it stays for 30 days. The departure epoch $t_0$ is allowed to slide (prior value 61880.6 MJD), and the total time of flight is fixed at 981.1 days. The thrust is 0.3~N, the specific impulse 3000~s, and the initial mass 2000~kg; the objective is minimum fuel consumption. Backend model: DeepSeek-V4-Flash.

\textbf{Stage 1---constraint parsing.} The ConstraintParser correctly extracts 6 \texttt{interior\_constraints} (3 rendezvous + 3 coast) and identifies \texttt{t0\_prior} and \texttt{total\_dt}. The parser splits each STAY constraint into two independent entries---a rendezvous (position/velocity match at the arrival epoch) and a coast (position/velocity match at the end of the stay)---and the LLM must merge them into a single STAY structure during the subsequent derivation.

\textbf{Stage 2---PMP derivation.} The PMPDeriver outputs the complete indirect formulation:

\begin{itemize}
  \item \textbf{Hamiltonian:}
  \begin{equation}
    H = \boldsymbol{\lambda}_r \cdot \mathbf{v} - \boldsymbol{\lambda}_v \cdot \frac{\mu \mathbf{r}}{r^{3}} + T u \left( \frac{\lambda_0}{I_{sp}\, g_0} - \frac{\lVert \boldsymbol{\lambda}_v \rVert}{m} - \frac{\lambda_m}{I_{sp}\, g_0} \right);
    \label{eq:s07ham}
  \end{equation}
  \item \textbf{Switching function:}
  \begin{equation}
    S = \frac{\lambda_0}{I_{sp}\, g_0} - \frac{\lVert \boldsymbol{\lambda}_v \rVert}{m} - \frac{\lambda_m}{I_{sp}\, g_0},
    \qquad u = \begin{cases} 1, & S > \varepsilon, \\ 0, & S < -\varepsilon; \end{cases}
    \label{eq:switch}
  \end{equation}
  \item \textbf{Variable structure} (48 dimensions): $\lambda_0$ (1) + initial costates $\boldsymbol{\lambda}_{r0}, \boldsymbol{\lambda}_{v0}, \lambda_{m0}$ (7) + $3\times$ STAY multiplier groups (each of size 13 = 6 arrival + 6 departure + 1 time) + $t_0$ offset (1) = 48;
  \item \textbf{Shooting equations} (48 dimensions): terminal rendezvous $\mathbf{r}(t_f) - \mathbf{r}_{\text{target}}$ (3) + $\mathbf{v}(t_f) - \mathbf{v}_{\text{target}}$ (3) + $\lambda_m(t_f) = 0$ (1) + $3\times$ STAY (each of size 13 = 6 arrival position/velocity matches + 6 costate continuities + 1 $H$-jump) + $H(t_0) = 0$ (1) + spherical normalization
  \begin{equation}
    \lVert \texttt{sphere}(13) \rVert^{2} - 1 = 0
    \label{eq:sphere}
  \end{equation}
  (1) = 48.
\end{itemize}

A noteworthy derivation result: the LLM correctly recognizes that each STAY constraint produces 6 arrival multipliers and 6 departure multipliers (12 in total), with the departure multipliers entering \texttt{Y0} with a negative sign (\texttt{Y0[7] = -dep\_var}), reflecting the costate jump across the stay arc. The $H$-jump equation contains the target-body velocity terms of both the arrival and departure endpoints:
\begin{equation}
  H(t_k^{-}) - H(t_k^{+}) - \boldsymbol{\nu}_{\text{arr}} \cdot \mathbf{v}_{\text{target}} - \boldsymbol{\nu}_{\text{dep}} \cdot \mathbf{v}_{\text{target}} = 0.
  \label{eq:hjump}
\end{equation}

\textbf{Stage 3---SymPy verification.} The symbolic verification script completes in 13 seconds and returns PASSED. The symbolic partial derivatives of the costate ODEs agree with the LLM's derivation.

\textbf{Stage 4---code generation.} The CodeGenerator produces 21{,}511 characters of C++ code in 72 seconds. The generated \texttt{objective\_RK} function correctly implements: (1) conversion of time variables to actual MJD via $\text{prior} \pm \text{range}$; (2) resetting the state components of \texttt{Y0} with the asteroids' exact positions and velocities at each rendezvous epoch after each integration leg (the multiple-shooting idea); (3) propagating the stay arcs with \texttt{fnull} (zero thrust) for 30 days; and (4) evaluating $H(t_0)$ before propagation.

\textbf{Stage 5---compilation and solving.} g++ compiles without errors. The solver reports:

\begin{itemize}
  \item 48 variables, up to 1000 random restarts;
  \item convergence on the very first restart ($\text{best\_res} = 9.1\times10^{-10}$);
  \item $\varepsilon$-homotopy descending from 1.0 to 0.0 in 46 steps, final residual $6.18\times10^{-14}$;
  \item final mass $m_f = 1522.46$~kg, equivalent $\Delta v = 8026.4$~m/s;
  \item terminal position error 73.3~m, velocity error $1.07\times10^{-5}$~m/s.
\end{itemize}

The structure of the 48-component solution vector is summarized in the table below.

\begin{table}[htbp]
  \centering
  \caption*{S07 solution vector structure (48 components).}
  \begin{tabular}{@{}lll@{}}
    \toprule
    Component & Physical meaning & Value \\
    \midrule
    x[0]      & $\lambda_0$                          & 0.836 \\
    x[1--3]   & $\boldsymbol{\lambda}_r(t_0)$        & $(-0.194,\ -0.317,\ 0.112)$ \\
    x[4--6]   & $\boldsymbol{\lambda}_v(t_0)$        & $(-0.069,\ -0.058,\ -0.036)$ \\
    x[7]      & $\lambda_m(t_0)$                     & 0.375 \\
    x[8--13]  & STAY1 arrival multipliers            & $(-0.155,\ -0.320,\ -0.033,\ 0.003,\ 0.080,\ -0.058)$ \\
    x[14--19] & STAY1 departure multipliers          & $(0.242,\ -0.128,\ -0.238,\ 0.092,\ -0.031,\ -0.020)$ \\
    x[20]     & STAY1 epoch offset                   & $-0.298$ \\
    x[21--33] & STAY2 multiplier group + epoch       & $\cdots$ \\
    x[34--46] & STAY3 multiplier group + epoch       & $\cdots$ \\
    x[47]     & $t_0$ offset                         & $-1.867$ \\
    \bottomrule
  \end{tabular}
\end{table}

All 48 residual components are below $10^{-13}$, verifying the completeness and correctness of the shooting system.

\begin{figure}[htbp]
  \centering
  \includegraphics[width=0.85\textwidth]{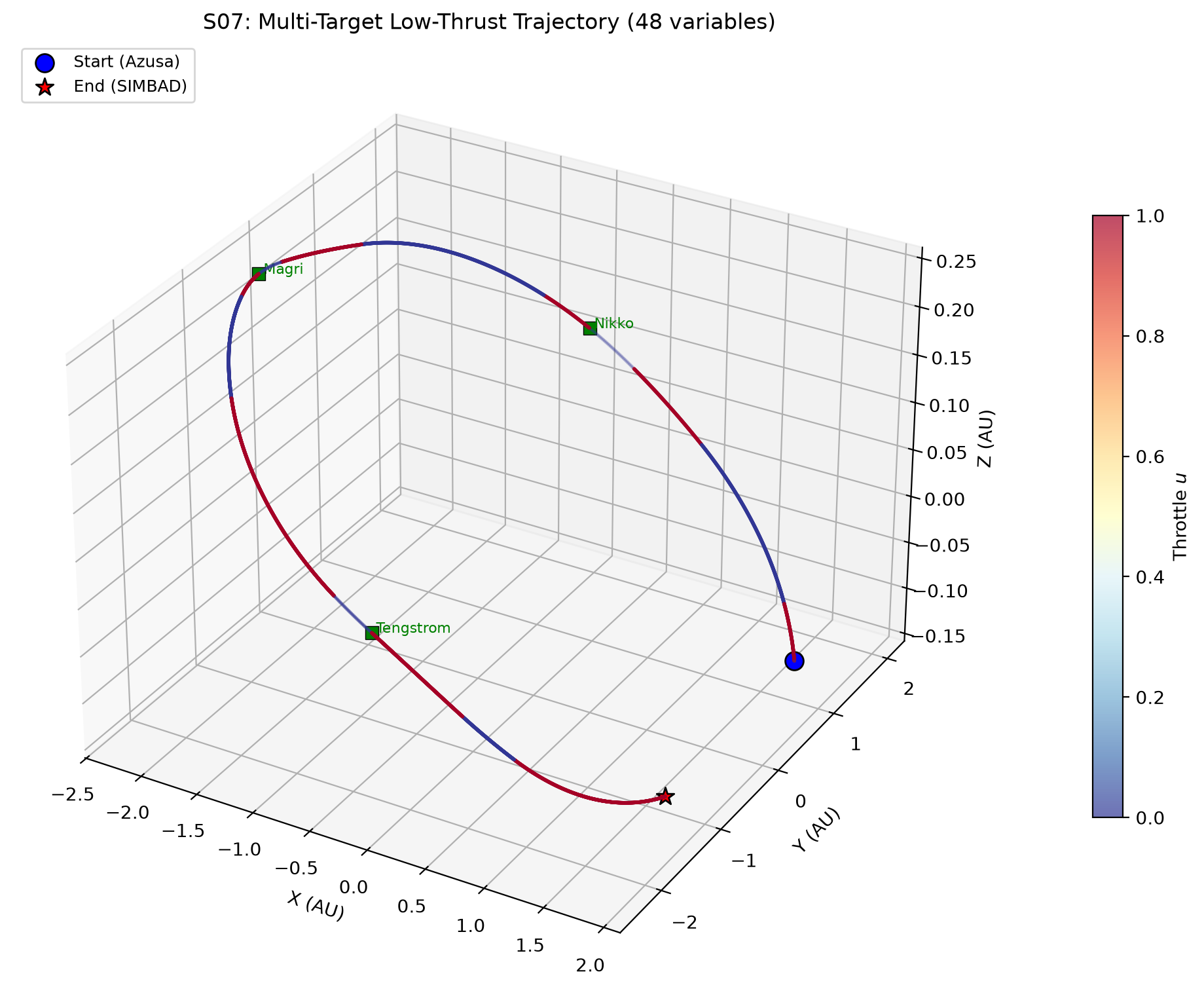}
  \caption{S07 three-dimensional trajectory (48-variable converged solution). Color encodes the throttle magnitude $u$; green squares mark the stay points, the blue circle the departure asteroid Azusa, and the red star the arrival asteroid SIMBAD.}
  \label{fig:s07traj}
\end{figure}

\textbf{Trajectory visualization.} Figure~\ref{fig:s07traj} shows the three-dimensional trajectory of the converged solution, color-coded by the throttle magnitude $u$. The spacecraft departs from Azusa, passes through the three stay points Nikko, Magri, and Tengstrom (green squares), and finally arrives at SIMBAD. The trajectory clearly exhibits the bang-bang control structure---thrust arcs (red) alternate with coast arcs (blue).

\begin{figure}[htbp]
  \centering
  \includegraphics[width=0.85\textwidth]{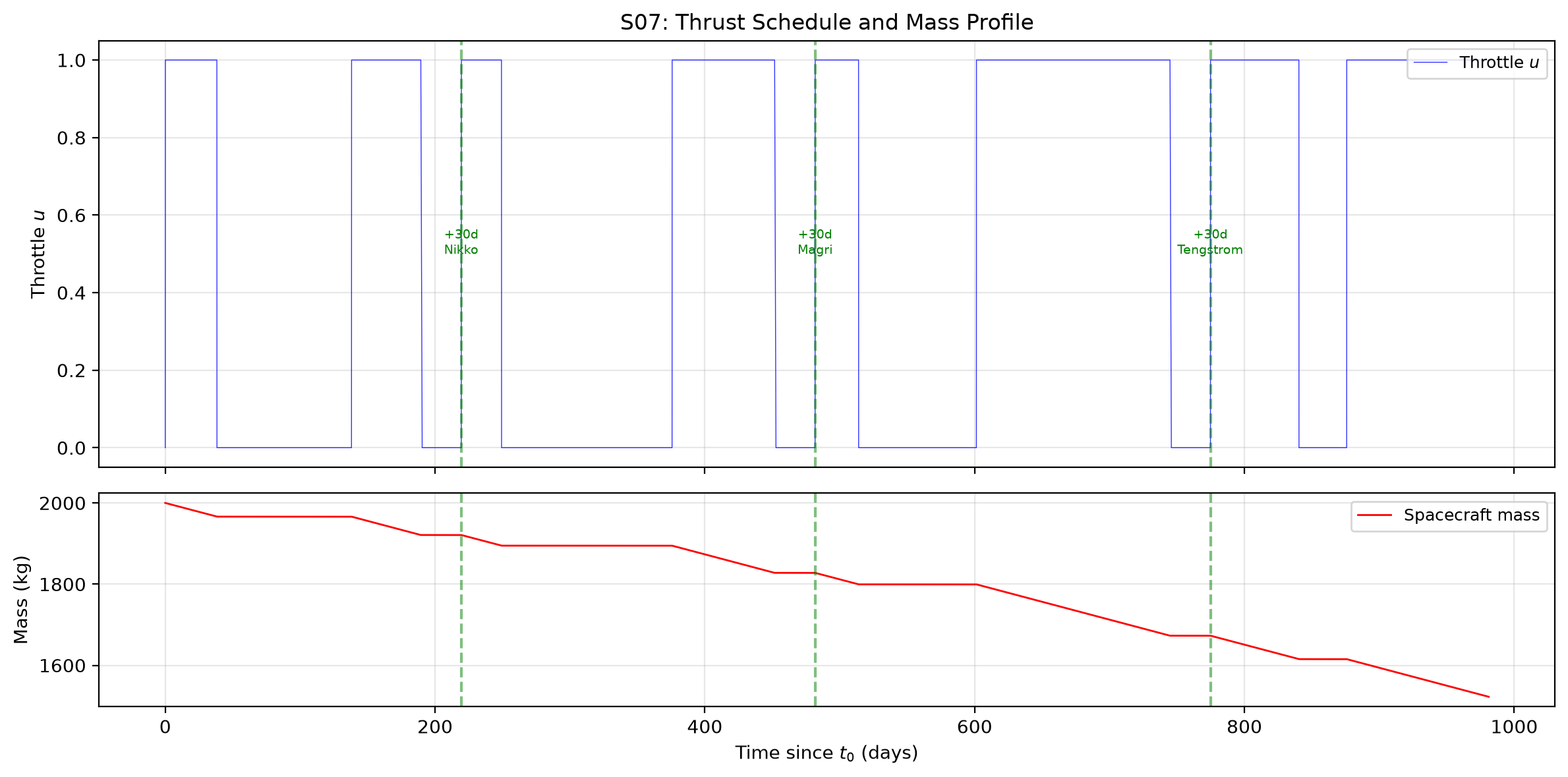}
  \caption{S07 thrust profile and mass history. Upper: throttle $u(t)$, with green dashed lines marking the stay arcs; lower: spacecraft mass $m(t)$.}
  \label{fig:s07thrust}
\end{figure}

Figure~\ref{fig:s07thrust} shows the thrust profile and mass history. The upper panel is the throttle $u(t)$, exhibiting the typical bang-bang structure ($u = 0$ or $u = 1$) with smoothed transition regions from the homotopy; the three green dashed lines mark the stay arcs. The lower panel shows the spacecraft mass decreasing monotonically from 2000~kg to 1522.46~kg, i.e., a propellant consumption of 477.54~kg (equivalent $\Delta v = 8026.4$~m/s). The thrust-on fraction is about 60.4\%, indicating that the optimal solution coasts for roughly 40\% of the time.

This case study shows that HELIOS can autonomously handle complex optimal control problems involving stay arcs, free epochs, and multiple interior constraints, without any human intervention.

\section{Discussion}

\subsection{System Adaptability}

The key advantage of HELIOS is its \emph{general adaptability to constraint types, dynamics models, and optimization objectives}:

\textbf{Constraint adaptability.} Through the unified framework of Rule~B2, the system automatically handles constraints with free parameters (the $\delta$ and $\phi$ of a GA) without per-constraint derivations. In traditional practice these must be derived by the researcher one by one, whereas HELIOS only needs the physical form of the constraint stated in the problem description.

\textbf{Dynamics adaptability.} Through the four-module code-generation architecture, the system produces complete custom code for different dynamics models. From chemical thrust (bang-bang) to solar sail (continuous cone angle), one only needs to state the dynamics equations in the problem description, and the LLM derives the Hamiltonian structure and the optimal control law autonomously.

\textbf{Objective adaptability.} Minimum-fuel problems (Lagrange form, $L = T u / (I_{sp}\, g_0)$) and minimum-time problems (Mayer form, $\varphi = t_f$) have different Hamiltonian structures:
\begin{equation}
  J_{\text{fuel}} = \int_{t_0}^{t_f} \frac{T\,u}{I_{sp}\, g_0} \, \mathrm{d}t,
  \label{eq:jfuel}
\end{equation}
\begin{equation}
  J_{\text{time}} = t_f .
  \label{eq:jtime}
\end{equation}
The unified PMP derivation framework handles both automatically.

\subsection{Prompt Engineering and Pipeline Improvements}

The derivation capability of the system originates mainly from the general rule set embedded in the prompt. These rules are not problem-specific ``cheat'' hints but general knowledge distilled from PMP theory:

\begin{itemize}
  \item Rule~B2 follows from the general theory of PMP: any free parameter in a constraint requires a stationarity condition;
  \item Rules G1--G4 follow from basic properties of the minimum principle;
  \item Rules G5--G6 forbid placeholders and simplifications in the derivation, requiring the full $3\times3$ Jacobian for the $J_2$ costate gradient;
  \item the ``Common Mistakes'' entries are error-prone points discovered in practice, but every one of them is general in nature.
\end{itemize}

Beyond the prompt, three key pipeline improvements significantly enhance robustness:

\textbf{Compilation repair loop.} On compilation failure, the compiler error messages are fed back to the LLM to repair the code (at most 3 rounds). This mechanism raises the compilation success rate from 36\%--55\% to 64\%--100\%; the repaired issues are pure code-quality problems (duplicated definitions, signature mismatches, etc.) that involve no change to the mathematical logic.

\textbf{SymPy verification in advisory mode.} SymPy verification failures no longer block the pipeline but serve as advisory checks. Code generation continues after a failed verification, avoiding pipeline crashes caused by false negatives (equivalent expressions not recognized). Statistics show that the SymPy verifier achieves 75\% accuracy, with a 25\% error rate.

\textbf{STAY constraint grouping.} The parser splits each STAY constraint into two independent entries (rendezvous + coast), and the LLM must first recognize and merge them into a single STAY constraint before counting variables. The enforced grouping step (Step~0b-STAY) avoids the double-counting problem.

\subsection{Limitations}

\textbf{Numerical convergence.} The convergence of the shooting method is strongly affected by the problem size: as the number of shooting variables increases, the probability that a random initial guess falls inside the basin of convergence decays rapidly. For scenarios with 17+ variables (e.g., S10 gravity assist), the random-restart strategy often fails to converge within the allowed time. It should be noted that the current evaluation relies on random initial guesses, which is not the most effective initialization strategy; the convergence dimension (d4) therefore carries an element of chance, and the corresponding scores should be interpreted with appropriate caution. A systematic study of initialization techniques---including physics-informed initial-guess construction, continuation from reduced problems, and warm-start from approximate solutions---is a necessary next step to make the convergence evaluation more robust and reproducible.

\textbf{SymPy verification.} For complex dynamics models (e.g., solar sail), symbolic verification times out due to expression explosion. More efficient verification strategies are needed (e.g., verifying only the gravitational part, or increasing the timeout).

\textbf{LLM output nondeterminism.} Two inference runs on the same problem may produce code with slightly different styles. For production-grade applications, deterministic post-processing is recommended (e.g., checking the linear independence of the shooting equations).

\subsection{Future Directions}

\begin{enumerate}
  \item \textbf{Initialization strategy:} systematic initial-guess construction algorithms (physics-informed, continuation from reduced problems) and warm-start techniques to replace pure random guessing, improving convergence robustness and reproducibility;
  \item \textbf{Differential-evolution global search:} use the DE algorithm of the \texttt{halo} library for global search at the $\varepsilon = 1$ stage;
  \item \textbf{SymPy verification optimization:} for complex dynamics models, verify only the gravitational part of the costate equations;
  \item \textbf{More dynamics models:} solar radiation pressure, atmospheric drag, and higher-fidelity gravity field models;
\end{enumerate}

\section{Conclusion}

This paper has presented HELIOS, an LLM-based autonomous indirect trajectory optimization agent. The core innovations of the system are:

\begin{enumerate}
  \item a \textbf{constraint-adaptive derivation framework} that automatically handles arbitrary constraint combinations through the unified $\psi(\mathbf{x}, \mathbf{p}) = 0$ form and Category~C free-parameter identification;
  \item \textbf{dynamics-adaptive code generation}, in which the four-module architecture supports complete code generation for non-standard dynamics models;
  \item a \textbf{general derivation rule set} (Rules G1--G4, B2, etc.) distilled from PMP theory rather than tailored to specific problems.
\end{enumerate}

Experiments validate the system on 11 scenarios: the compilation success rate reaches 100\% (11/11) at best, and multiple scenarios converge to optimal solutions consistent with reference solutions, including a 48-variable STAY multiple-shooting problem and a 17-variable gravity-assist problem. The multi-model comparison (8 LLMs, total scores 250--905) verifies the model-agnostic nature of the architecture and reveals a positive correlation between model scale and derivation capability.

HELIOS demonstrates the value of LLMs for symbolic derivation in astrodynamics---compressing a manual derivation process that traditionally takes days into an automated pipeline completed in minutes, while maintaining general adaptability across constraint types, dynamics models, and optimization objectives.

\section*{Declaration of AI Assistance}
During the preparation of this work, the authors used large language models for code generation and debugging assistance in the development of the HELIOS system, and for translation and language polishing of the manuscript. The authors reviewed and verified all AI-generated content and take full responsibility for the integrity of this publication.

\end{document}